\documentclass[conference]{IEEEtran}
\IEEEoverridecommandlockouts
\usepackage{cite}
\usepackage{amsmath,amssymb,amsfonts}
\usepackage{algorithmic}
\usepackage{graphicx}
\usepackage{textcomp}
\usepackage{xcolor}
\def\BibTeX{{\rm B\kern-.05em{\sc i\kern-.025em b}\kern-.08em
    T\kern-.1667em\lower.7ex\hbox{E}\kern-.125emX}}
\begin{document}

\title{Music Sentiment Transfer\\
{\footnotesize}
\thanks{This research was funded by the National Science Foundation.}
}

\author{\IEEEauthorblockN{Miles Sigel}
\IEEEauthorblockA{\textit{Department of Computer Science} \\
\textit{Rice University}\\
Houston, TX, USA \\
msigel@rice.edu}
\and
\IEEEauthorblockN{Michael Zhou}
\IEEEauthorblockA{\textit{Department of Computer Science} \\
\textit{Cornell University}\\
Ithaca, NY, USA \\
mgz27@cornell.edu}
\and
\IEEEauthorblockN{Jiebo Luo}
\IEEEauthorblockA{\textit{Department of Computer Science} \\
\textit{University of Rochester}\\
Rochester, NY, USA \\
jluo@cs.rochester.edu}
}

\maketitle

\begin{abstract}
Music sentiment transfer is a completely novel task. Sentiment transfer is a natural evolution of the heavily-studied style transfer task, as sentiment transfer is rooted in applying the sentiment of a source to be the new sentiment for a target piece of media; yet compared to style transfer, sentiment transfer has been only scantily studied on images. Music sentiment transfer attempts to apply the high level objective of sentiment transfer to the domain of music. We propose CycleGAN to bridge disparate domains. In order to use the network, we choose to use symbolic, MIDI, data as the music format. Through the use of a cycle consistency loss, we are able to create one-to-one mappings that preserve the content and realism of the source data. Results and literature suggest that the task of music sentiment transfer is more difficult than image sentiment transfer because of the temporal characteristics of music and lack of existing datasets.
\end{abstract}


\section{Introduction}
Sentiment transfer is a relatively inchoate topic. In contrast to well-known extant tasks such as style transfer, sentiment transfer focuses on modifying the high-level features of the source data to change its overall mood to people; for instance, an original music piece can be transferred to a more positive vibe to give people a feeling of enjoyment, without having to modify the overall melodic structure. More concretely, sentiment transfer modifies an input data from a source sentiment into a target sentiment by learning the underlying structures of different sentiment classes while preserving the original content. 

Unlike the well-studied style transfer tasks, sentiment transfer has only been studied in images. Reference \cite{IST} proposes a Sentiment-aware GAN (SentiGAN) to perform object-level sentiment transfer for each contained local object in an image, transferring the sentiment by changing color-based features (e.g. brightness, saturation, dominant color, contrast) of the objects while preserving the overall content such as the object contours and textures of the input image. Reference \cite{Global IST} transfers the overall sentiment of the entire image by using feature maps from a pre-trained CNN.

Music and audio sentiment transfer, on the other hand, do not contain as well-defined features as images. The structural natures of audio and music are completely different from images; specifically, audio uses the time domain, and music contains unique structural attributes such as key, tempo, time signature, and dynamics. Furthermore, in raw audio, there is a higher tendency for noise and lack of structure due to high sampling rates. 

Music sentiment transfer is a completely unexplored task. Considering the inability of music to be separated into a clear content and style code like in image style transfer, domain transfer networks such as CycleGAN are common in transfer tasks involving dissimilar domains, including music. Reference \cite{Symbolic Music CycleGAN} is the first work to explore music genre transfer in symbolic audio data. In this paper, we attempt to utilize the CycleGAN network on a novel dataset containing binary sentiment labels. In this task, we attempt to transfer music from a negative sentiment domain to a positive sentiment domain and vice versa. 

Our contributions are summarized as follows:
\begin{itemize}
    \item We are the first to explore music sentiment transfer. 
    \item We use a CycleGAN implementation as the core component for music sentiment transfer. 
    \item We create a single-track symbolic music dataset with labeled sentiments by transforming sentiment-labelled video game piano soundtracks into \textit{piano roll arrays}, which represent note-controlled data. 
\end{itemize}

\begin{figure}
    \centering
    \includegraphics[width = 0.3 \textwidth]{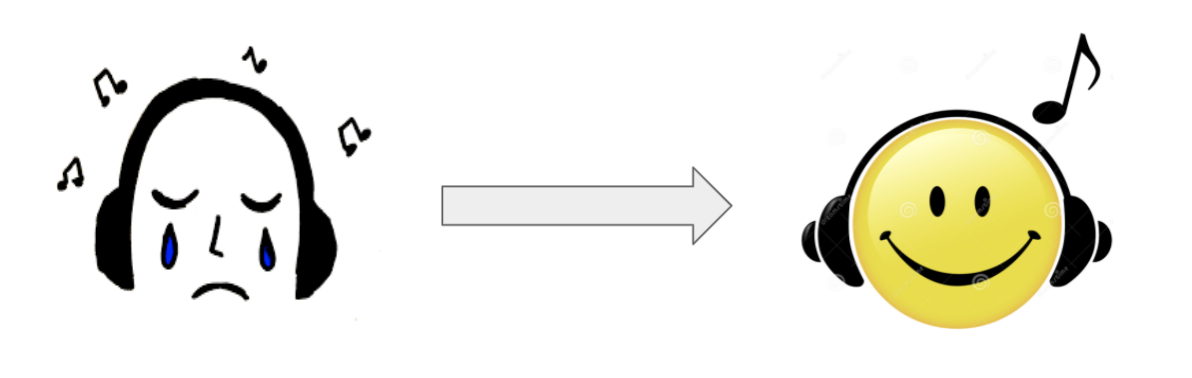}
    \label{fig:my_label}
\end{figure}

\section{Related Works}

The most well-known transfer task is style transfer. Style transfer is a topic in computer vision pertaining to transferring aspects of the underlying characteristics of one media to another. This task, originally studied in images and introduced in \cite{Gatys}, applied the style characteristics (ex. color, texture, brush strokes) of a source image to a target image through an iterative process that minimized the distance between “style characteristics.” Specifically, the research partitioned images into a “style code” and a “content code” which allowed for the disentanglement of style from an image. Using the style code of the source image, style transfer was possible without completely distorting the underlying content of a target image. Style transfer grew quickly after its advent, as researchers started to find ways to not only optimize this process but also use them in both academic and industrial applications  \cite{Huang_2017_ICCV} \cite{li2017universal} \cite{neural_style_tranfer_review}. Further, image based style transfer techniques were subsequently applied to video through very similar methods of matching style statistics within images \cite{huang2017real}.

Building upon style transfer techniques, \cite{IST} extended the framework to build a SentiGAN, a model that transfers the sentiment of one image to another. Moreover, sentiment transfer is not accomplished by artistic style transfer because of the importance of realism in sentiment transfer in natural images, warranting the need for parameters controlling the strength of content features in the media getting transformed. Modifying a higher level aspect, sentiment, means modifying perceptual features that don't directly distort the identity of the image, something that existing style transfer cannot accomplish. 

In the domain of audio, style transfer has taken form based on the different types of audio tasks being studied. For example, in the domain of natural language, voice conversion networks are used to change a voice from male to female \cite{lee}. To complete voice conversion tasks, raw audio is commonly used and a mapping is learned between the underlying frequencies. Further, the field of audio style transfer is greatly limited by the lack of paired data, meaning that general frequency conversion is possible but a fine grained mapping is difficult. Most hindered by such a problem is that of this paper, music style transfer. The domain of music style transfer brings into question the foundation of style in music. Contrary to image style transfer with clear pixel level style characteristics (color, brightness), music incorporates not only the time-domain but also different instrumentation and musical clef differencers. As a consequence of such limitations, audio research has gravitated toward the symbolic form of music commonly used today, MIDI (Musical Instrument Digital Interface) \cite{kotecha}.

Music sentiment transfer is an unexplored task, in which the sentiment classification of a piece of music is applied to another piece of music, transforming its sentiment. Considering the inability of music to be separated into a clear content and style code, domain transfer networks such as CycleGAN are common in transfer tasks involving music due to the domains being dissimilar from one another. Another area of exploration in the domain of sentiment in music is in the ability for a neural network to generate sentiment aware music \cite{Ferreira}. Work done in music emotion classification networks have advanced in recent years, using various data formats besides raw audio signals such as MFCC features and log-mel spectrograms \cite{HIZLISOY2021760} \cite{sarkar2020recognition}. 

\section{Methodology}

\subsection{Data Representation: Symbolic vs Raw Audio}

We considered two types of representations for music data: \textit{raw audio data} (e.g. WAV, MP3) and \textit{symbolic music data} (e.g. MIDI). 

We first attempted to use raw audio data. Specifically, there is existing research on Voice Conversion GAN networks which we experimented on with our novel piano datasets, thinking that a network that could learn the difference between male and female voice could learn the two different voices of a depressed piano and happy piano \cite{pasini2019melgan} \cite{MaskCycleGAN-VC} \cite{CycleGAN-VC3} \cite{StarGAN-VC2} \cite{lee}. After tuning hyperparameters in \cite{pasini2019melgan}, we found that this network couldn't learn the deeper structure necessary for sentiment transfer, and could only learn relationships between different frequency ranges or families, namely male and female. Moreover, we concluded that these types of networks were unable to learn a comprehensive mapping between sentiments, meaning we would need to simplify the data representation or expand the network. We choose to simplify the data. 

Instead, we chose to use symbolic data, aka MIDI data, a discrete music format similar to sheet music. MIDI (Musical Instrument Digital Interface) format encodes essential music structure properties, such as how long a note is held, key, time signature, loudness of notes, etc., which can be associated to music sentiments. For example, a melancholy song can have soft dynamics and a minor key, while a jovial song can have a full, rich tone, louder dynamics, and a major key. Additionally, we were able to reduce the dimensionality of the data by only using binary note-presence values. In effect, MIDI data could be broken down into a simple matrix with time and binary note on/off values, commonly referred to as a piano roll. 

\subsection{Approaches: Data-Driven vs Knowledge-Driven}

We considered two different approaches for the musical sentiment transfer task: the \textit{data-driven} approach, which uses machine learning to learn the underlying patterns of music, and transfer the sentiment based on these underlying patterns; the \textit{knowledge-driven} approach, which uses music theory to change certain aspects of musical structure of the original piece (e.g. minor to major keys for negative to positive transfer). The data-driven approach requires a large amount of extant data to train the models accurately, while the knowledge-driven approach simply requires expert knowledge of music theory to manipulate the original content structure and data does not have to be present. 

Since it is difficult to find existing symbolic music datasets with sentiment labels, especially for single-track data, we decided to create our own dataset by transforming symbolic music data from VGMIDI \footnote{https://github.com/lucasnfe/vgmidi} into piano roll arrays \cite{Ferreira}. In addition to machine learning models, future work should also incorporate expert knowledge of music theory to provide insight between musical structures and sentiments. 

\subsection{Model Architecture}

We use a CycleGAN implementation from \cite{Symbolic Music CycleGAN} to bridge the two domains of music sentiment - happy and sad. The goal of a CycleGAN is to learn a one-to-one mapping between two domains with unlabeled data. Like other GANs, the CycleGAN consists of generators which try to generate fake data, and discriminators that tell whether the generated data is real or fake. The goal is to have the generators fool the discriminator, and generate realistic images in the target domain. The CycleGAN in particular allows transferring images back and forth between domains while preserving the content information. We used the model introduced by\cite{Symbolic Music CycleGAN} which is a variant of the CycleGAN, however, contains an additional discriminator that attempts to train the generator to retain basic music structure. The loss is accomplished by comparing \(\Tilde{x}\) to a larger sample of music.

\begin{figure}
    \centering
    \includegraphics[width = 0.5 \textwidth]{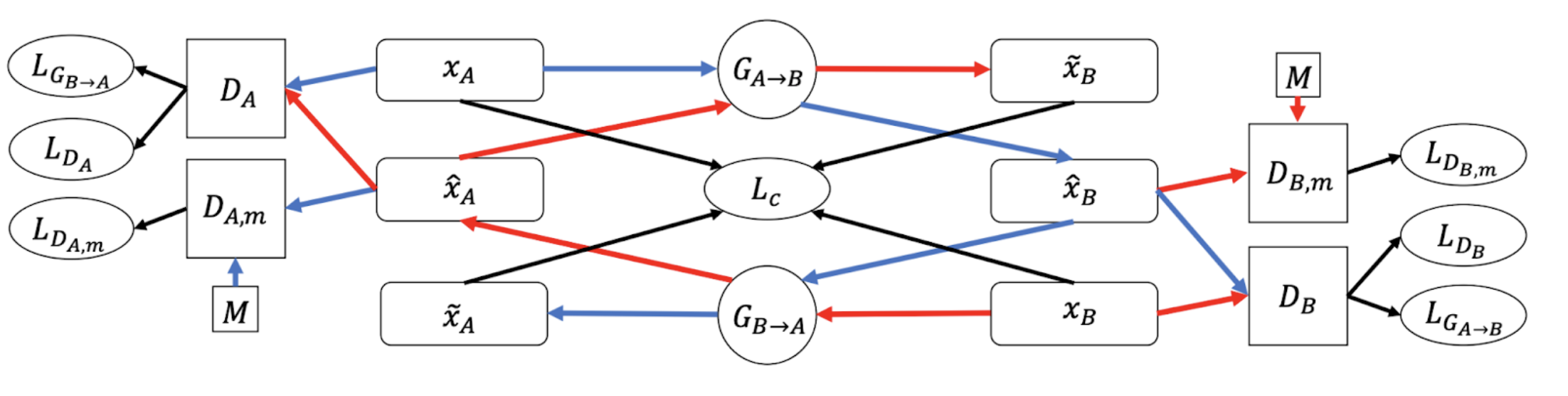}
    \caption{CycleGAN architecture as proposed in \cite{Symbolic Music CycleGAN}.}
    \label{fig:my_label}
\end{figure}

\section{Dataset}

Sentiment labeled music datasets are relatively new and slowly being introduced into the research. One such MIDI labeled dataset was constructed by \cite{Ferreira} and drew from a pool of 64-bit video game tracks \cite{Ferreira}. Using Amazon Mechanical Turk, \cite{Ferreira} had 30 people hand annotate the tracks with their valence and arousal estimates as well as creating the tool that allowed for the human annotation. For this project, we utilized the dataset VGMIDI dataset by creating a program to convert the multitrack MIDI data into binary piano rolls. Following the specifications in \cite{Symbolic Music CycleGAN}, the data was re-sampled to a maximum resolution of 16th beats and constrained to represent the 84 MIDI notes played on a piano, resulting in a (1, 64, 84) piano roll matrix. While transcription between raw audio to MIDI is possible using CNNs, no comprehensive dataset has been created on the topic. During the process of research, tools were created to make datasets between raw audio and spectrograms as well from MIDI to piano rolls, both of which can be found in the project Github \footnote{https://github.com/milesigel/Audio-Sentiment-Transfer}. 

\section{Experimental Setup}

The CycleGAN network created by \cite{Symbolic Music CycleGAN} was used as the experimental network, besides the fact that we utilized a PyTorch version. For our dataset, we used a dataset adapted from the VGMIDI dataset. To prepare the data, we transformed 203 multitrack piano rolls into binary piano rolls. We decided to normalize the key signature to 4/4, removing tracks with a ¾ time-signature, producing 153 viable MIDI files.  Using the pypianoroll and pretty-midi packages, we created (1, 64, 84) piano rolls and labelled them according to their valence values into binary positive and negative categories. After downsampling to contain equal classes, 2631 samples were in each class. Additionally, in the dataset sampled by the additional discriminator in the network, a joined folder of both binary classes was created. The code was run for 150 epochs or until the generators and discriminators converged. 

The code resulted in output, but the network output was too small to be saved and interpreted by a MIDI interpreter. As a result, we are continuing to run experiments with different network versions in Tensorflow and Pytorch and will update results as they emerge.

\section{Results}

The results of this study were compiled using a Pytorch implementation \footnote{https://githubmemory.com/repo/Asthestarsfalll/Symbolic-Music-Genre-Transfer-with-CycleGAN-for-pytorch} of the CycleGAN framework first researched in the paper Symbolic Music Genre Transfer with CycleGAN from \cite{Symbolic Music CycleGAN}. Using the dataset created by our research team, we received outputs from running the test code on the network. Even though our data conformed to the specified data representation, the network produced outputs that were not interpretable by the MIDI converter, indicating a problem with the network output. 

We then tried out the Tensorflow implementation \footnote{https://github.com/sumuzhao/CycleGAN-Music-Style-Transfer-Refactorization} of the CycleGAN in the same paper. We are still awaiting results from this model implementation.

\section{Discussion}

\subsection{Datasets}

The most challenging aspect of this research is the lack of a sentiment labeled dataset. In this experiment, we don’t have access to paired data, so we needed to create our own dataset. We attempted to use existing MIDI sentiment labeled datasets, but due to the need for single track audio, we had to create our own. In order to create our own, we transformed MIDI files to their respective piano roll arrays. For future work, we suggest that a clearly annotated dataset be created for both symbolic and raw audio in a single track. Additionally, a paired dataset would allow for a more direct mapping between the sentiment domains. The data representation will continue to be a debate within this task, however, we believe that symbolic audio will likely be the main format seen going into the future. For our task, we attempted to complete the sentiment transfer on single-track, single-instrument MIDI files. As aforementioned, we utilized the VGMIDI dataset, video game soundtracks that could be played on a single piano. Piano Roll representations are able to capture MIDI data into a format that is easily interpretable by current networks. More dimensions could be added to pianorolls to include additional parameters such as note velocity. We expect as this task progresses, making more comprehensive MIDI datasets will allow for the exploration into the effect of different characteristics of music on sentiment and style. 

\subsection{Models}

Sentiment transfer for music is different then the approach used for images due to the differing data representations. Music, once turned into discrete piano rolls, could not be separated into content and style codes. Therefore, a network like CycleGAN that uses unpaired data and can work with arbitrary data representations is an understandable choice \cite{CycleGAN}. Autoregressive generative networks could potentially be used to accomplish this task but might be hindered by the ability to preserve source content \cite{maduskar}. MelGAN-VC is another network similar to CycleGAN that uses a siamese network to preserve the distance between classes in a latent space \cite{MelGAN-VC}. Models using spectrograms introduce noise, so many extant networks used for images are not congenial. Transfer models for symbolic music look to be still in their infancy, we believe mainly due to the lack of datasets and choice of representation into the models. GAN based networks are commonly used for transfer tasks in images and warrant their use in audio. The network by \cite{Symbolic Cycle GAN} uses an additional loss that backprogopates information needed to retain the core elements of music (tempo regularity, time signature) in the generator. We believe that networks using music data will likely need to incorporate losses like this in order to retain underlying important trends in music composition and theory. Knowledge based approaches using parameter restraints and a deeper sense of musical encoding could attempt to design new data structures that allow for these features to remain embedded into the content code. 

\subsection{Data Representation}

Audio data comes in two main formats, symbolic data and raw audio data. Symbolic data is commonly represented in MIDI format, a discrete representation that annotates the notes and metadata of a piece of music. Using this format allows for less noise in the data compared to raw data, as the sampling rate of raw audio introduces many irregularities into the data. Additionally, for music, using raw data does not make some features of the underlying structure of the music apparent and relies on a network being able to learn long range temporal features. Using symbolic data, however, is potentially less applicable in real world situations which typically operate on raw audio.

\section{Conclusion}

Music and audio sentiment transfer is a completely novel task, but is an extension of the better-studied style transfer task. Music sentiment transfer, however, goes beyond just simply changing note tones, as sentiment is embedded in deeper structures of music. As there is a paucity of extant datasets that are labeled by sentiment for both raw audio and symbolic audio, we created our own dataset. While we propose a CycleGAN framework for generating the music samples with the specified sentiment, further research should combine both expert knowledge of music composition to tune parameters and provide insight into the relationship between musical structure and sentiment. We plan to also extend the task of sentiment transfer towards other modalities including raw speech audio and videos (with and without background sound).

\section*{Acknowledgments}

The authors would like to thank the National Science Foundation (NSF) for funding the Research Experience for Undergraduates program in which this work took place, the Kearns Center for Diversity at the University of Rochester for their part managing the REU experience at Rochester, the Georgen Institute for Data Science for organizing the \textit{Computational Methods for Understanding Music, Media, and Minds} REU, along with Wei Xiong and Prof. Zhiyao Duan of the Department of Computer Science at the University of Rochester for his support throughout this project and program.

\end{document}